\documentclass[lettersize,journal]{IEEEtran}
\usepackage{amsmath,amsfonts}
\usepackage{algorithmic}
\usepackage{algorithm}
\usepackage{array}
\usepackage{authblk}
\usepackage[caption=false,font=normalsize,labelfont=sf,textfont=sf]{subfig}
\usepackage{textcomp}
\usepackage{stfloats}
\usepackage{url}
\usepackage{verbatim}
\usepackage{graphicx}
\usepackage{xcolor}
\usepackage{multirow}
\usepackage{hyperref}
\usepackage{booktabs}
\usepackage{tabularx}
\usepackage{orcidlink}

\usepackage[switch]{lineno}

\begin{document}

\title{Cryogenic Materials Repository: A Public Resource and New Measurements for Cryogenic Research Applications}

\author{Henry E. Nachman\orcidlink{0000-0002-8694-1527}, Oorie Desai\orcidlink{0009-0000-9795-6209}, Nicholas Galitzki\orcidlink{0000-0001-7225-6679}, Daniel Lee\orcidlink{0009-0007-2928-7024}, JB Lloyd\orcidlink{0000-0003-1581-1626}, Tannishtha Nandi\orcidlink{0009-0003-9432-4056}, Ani Pagni\orcidlink{0009-0003-7911-7008}, Ray Radebaugh$^*$\orcidlink{0000-0002-9788-9736}, and Elle C. Shaw\orcidlink{0000-0001-5644-8750}
\thanks{Received 26 September 2025; revised 26 May 2026; accepted 06 July 2026; published 13 July 2026.  \textit{(Corresponding author: Henry E. Nachman.)} }
\thanks{Henry E. Nachman, Oorie Desai, Nicholas Galitzki, Daniel Lee, JB Lloyd, Tannishtha Nandi, Ani Pagni, and Elle C. Shaw are with the Department of Physics, The University of Texas at Austin, Weinberg Institute for Theoretical Physics, Texas Center for Cosmology and Astroparticle Physics, Austin, TX, USA. (e-mail: henry.nachman@utexas.edu)

$^*$Ray Radebaugh was a National Institute of Standards and Technology Fellow Emeritus. He passed away on February 11, 2026.}
\thanks{Contribution of NIST, not subject to copyright in the US.}}




\maketitle

\begin{abstract}
Low-temperature systems play a vital role in a variety of scientific research applications, including the next generation of cosmology and astrophysics telescopes. More ambitious cryogenic applications require precise estimates of the thermal conductivity of materials and thermal joints to meet project goals. We present the development of the Cryogenic Material Repository (CMR), a public GitHub repository of cryogenic material properties data created to support and enable researchers across scientific disciplines to accurately and efficiently design and assess cryogenic systems. We also present thermal conductivity measurements for a selection of carbon fiber reinforced polymer and aluminum alloy samples conducted with a Bluefors LD400 Dilution Refrigerator. These measurements, between $\mathbf{{\sim}0.07{-}2}$ K, are consistent with, and expand on existing sub-kelvin thermal conductivity data available on the repository.  

\end{abstract}

\begin{IEEEkeywords}
Aluminum, carbon fiber reinforced polymers (CFRP), cryogenic properties, data repository, thermal conductivity
\end{IEEEkeywords}
\vspace{-1em}
\section{Introduction} \label{sec:intro}

Low-temperature devices such as transition-edge sensors and kinetic induction detectors enable a host of increasingly sensitive research applications spanning the fields of experimental cosmology, particle physics, material science, quantum computing, and many more \cite{delucia_TransitionEdgeSensors_2024}\cite{austermann_millimeterwave_2018}. These devices rely on systems capable of operating at sub-kelvin temperatures. Designing, building, and modeling these systems requires knowledge of material properties across many orders of magnitude of temperature. Unifying the storage of available data can reduce design time and costs while improving outcomes across fields.

Past efforts to build a database of cryogenic material properties have compiled fits to properties such as thermal conductivity, specific heat, and coefficient of thermal expansion \cite{woodcraft_low_2009}.
A notable example is the website created by the Cryogenic Technologies Group at the National Institute of Standards and Technology (NIST)\cite{marquardt_cryogenic_2000}\cite{bradley_cryogenic_2006}\cite{bradley_properties_2016}. 
While the NIST website provides an easily accessible repository of property fits for a variety of materials, it lacks transparency in some of its source data, it cannot be easily expanded, and its format makes integration with other tools difficult. 
To address these concerns, we have developed the Cryogenic Materials Repository (CMR). The CMR is an easy-to-use, Python-based, public repository of thermal conductivity data, fits, and tools hosted on GitHub under the tag \verb+CMB-S4/Cryogenic_Material_Properties+. 

In Section \ref{sec:repo} we describe the CMR repository, its structure, the method for fitting thermal conductivity, and usage examples. Section \ref{sec:tc_testing} discusses the method for thermal conductivity testing of select Aluminum alloys and carbon fiber reinforced polymers (CFRP) at The University of Texas at Austin (UT Austin). Section \ref{sec:results} presents these new measurements in context with previous results.

\section{Materials Repository}\label{sec:repo}

The CMR contains thermal conductivity data, empirical and analytic fits to data, and additional thermal conductivity fits from the literature. The repository currently stores hundreds of datasets on experimentally measured thermal conductivity of more than 80 materials collected from published works spanning decades of research (Section \ref{sec:datarepo}). The data consists of many common materials used in low-temperature systems, such as aluminum and steel alloys, as well as custom cryogenic components (e.g., heat straps\footnote{Heat strap data is currently awaiting publication and will be added upon public release.}). The thermal conductivity fit parameters are compiled into downloadable comma-separated value (CSV) files, enabling integration with other codes or programs (Section \ref{sec:structure}). A robust Python-based fitting algorithm generates integrable functions of each material's thermal conductivity for export or use in thermal modeling (Section \ref{sec:fits}). Lastly, the CMR provides several examples and tools for using the repository to model a cryogenic system. The repository is designed to be a public tool that encourages transparency by ensuring all data is properly referenced and source code is documented and published (Section \ref{sec:tools}). The CMR is currently public, and further data and code contributions from the community are encouraged.

\vspace{-1em}
\subsection{Data Repository}\label{sec:datarepo}
The CMR stores datasets containing measurements of thermal conductivity as a function of temperature, $\kappa(T)$. Storing the thermal conductivity data enables the repository to compare different or conflicting measurements, produce its own fits, and ensure transparency by providing proper data source citations. Datasets are stored as individual CSV files within a subdirectory of the material folder and contain one column each for average temperature, thermal conductivity, and $\kappa/T$. Datasets are acquired directly from experimental labs and publications, or transcribed from published plots using online plot digitizing tools. 

\begin{figure}[ht]
    \centering
    \includegraphics[width=0.94\linewidth]{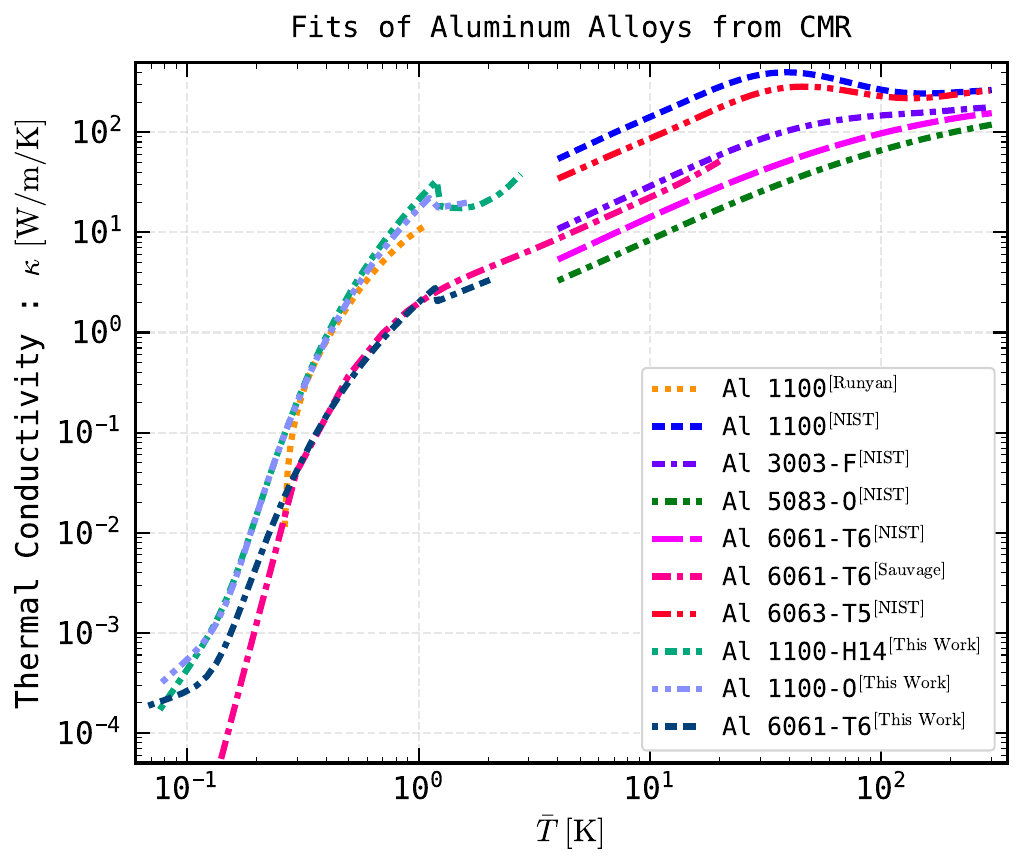}
    \caption{All available fits for the aluminum parent material stored on the CMR from room temperature to sub-$100$ mK. This plot and others available on the CMR can be used to compare different alloys. Some fits are made using the data collected as described in Section \ref{sec:fits}.}
    \label{fig:repo_alum}
\end{figure}

Materials are divided into specific alloys, material types, and components/devices, and each material is stored in a separate folder within the repository. Materials belonging to a category of parent materials are designated as such within the material's object-oriented attributes. An example of this is the parent material, \textit{Aluminum}, with child materials consisting of specific alloys, as shown in Figure \ref{fig:repo_alum}. This structure allows the data and fits to be attributed to the parent material just as with the child.

\vspace{-1em}
\subsection{Fit Types and Development}\label{sec:fits}

The repository contains both fits to thermal conductivity data and fits from published literature. The literature fits have been extracted from a variety of sources, relying largely on those collected on the NIST website\footnote{\href{https://trc.nist.gov/cryogenics/}{https://trc.nist.gov/cryogenics/}}. Literature fits are typically produced using a limited subset of the total data available for a material. The repository collects and compiles all the known published data and creates its own fits to maximize transparency. By default, the CMR makes no data quality cuts and uses all available datasets when creating a fit. Users can select datasets manually for a more customized experience. While many different fit types are cataloged in the repository, by default the data are fit to one of, or a combination of, two general types of fits: a polynomial function (\ref{eqn:nppoly}) and a polynomial log function (\ref{eqn:polylog}).

\noindent
\begin{tabularx}{\linewidth}{p{3.2cm}p{5.0cm}}
\begin{equation}\label{eqn:nppoly}
\hspace{-0.4cm}
  \kappa(T) = \sum_{n=0}^N a_n T^n;
\end{equation}
&
\begin{equation}\label{eqn:polylog}
  \hspace{-.4cm}
  \log_{10}(\kappa(T))=\sum_{n=0}^N a_n \log_{10}(T).
\end{equation}
\end{tabularx}

Fit functions must describe thermal conductivity across many orders of magnitude in temperature. Importantly, N must be less than the number of data points (usually less than 1/3 to 1/2). Additionally, the use of $\log$ functions helps prevent unrealistic weighting of the large values. Unfortunately, these polynomial fits cannot be extrapolated much beyond the fitted data. Because thermal conductivity usually varies rapidly with temperature, (\ref{eqn:polylog}) is often used in the NIST cryogenic material database to give equal weight to thermal conductivity over the entire temperature range of the fit \cite{marquardt_cryogenic_2000}. Another function option often used for fitting thermal conductivity data of metals below about 20 K is $\kappa / T$, which usually approaches a constant as $T\rightarrow 0$ and changes little below about 50 K. It also follows theoretical behavior and can be used to extrapolate beyond the range of data, including the millikelvin range.

Given the vast temperature range covered by many of the materials, a single fit is often unable to accurately describe thermal conductivity across the entire data range. Instead, the temperature range is often split into two or more sub-ranges, within which (\ref{eqn:nppoly}) or (\ref{eqn:polylog}) are used. The coefficients may be adjusted to force continuity of the functions at the boundaries.

\begin{figure}[ht]
    \centering
    \includegraphics[width=0.97\linewidth]{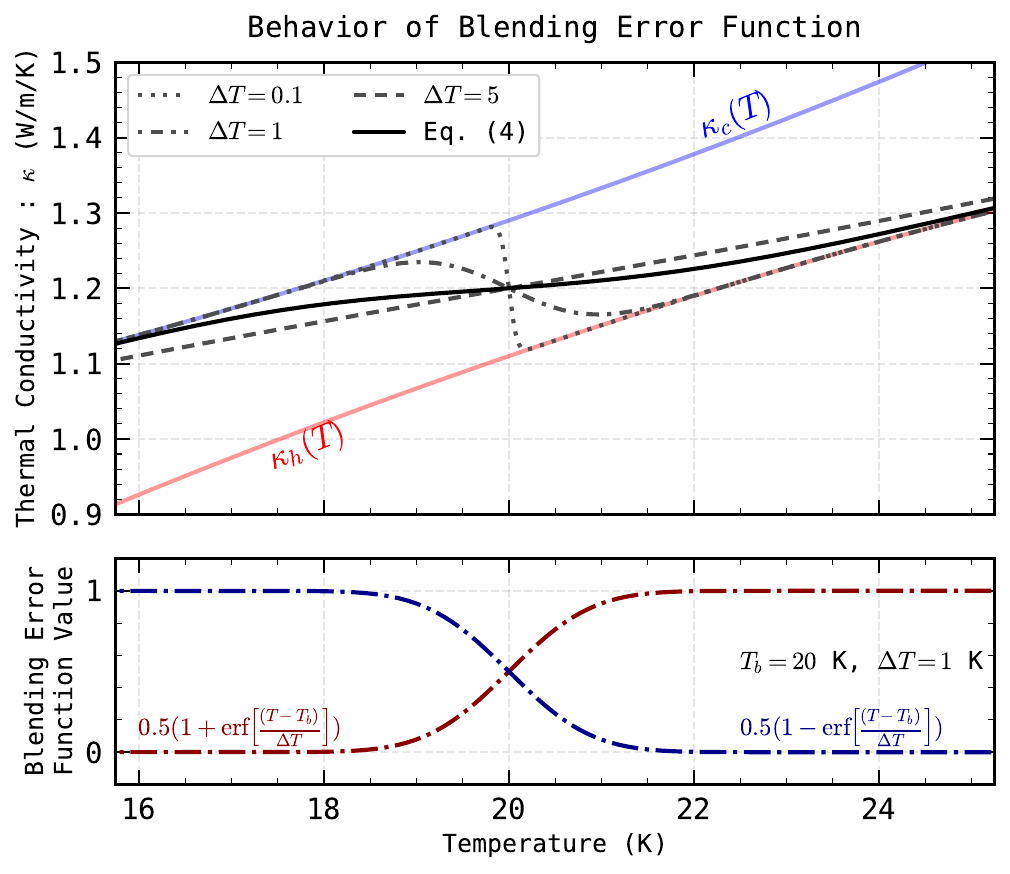}
    \caption{The bottom frame shows an example of the blending error function factors in (\ref{eqn:erf_theory}) with $T_b=20$ K and $\Delta T = 1$ K. The top frame shows their use in blending two mismatched functions for low, $\kappa_c(T)$, and high, $\kappa_h(T)$, regimes (exaggerated). Three gray lines show the resulting blend for different values of $\Delta T$. Blending with (\ref{eqn:erf_repo}) is shown as the solid black line.}
    \label{fig:erf_func}
    \vspace{-0.75em}
\end{figure}

A fitting procedure developed at NIST (see silicon thermal expansion on the NIST website \cite{cryogenic_technologies_group_nist_cryogenic_2009}) makes use of the error function as a factor to blend fits in adjacent regions while maintaining no discontinuities in the function or in any of its derivatives. This function takes the form of (\ref{eqn:erf_theory}) with the minus sign used as a factor for the lower region and the plus sign used as the factor for the upper region.

\begin{equation}
    0.5\cdot\left(1\pm\text{erf}\left[\left(T-T_b\right)/\Delta T\right]\right),
    \label{eqn:erf_theory}
\end{equation}
where $T_b$ is the blend or junction temperature and $\Delta T$ is the half-width of the blending region. For the NIST website, $T_b$ and $\Delta T$ are chosen with some trial and error to find the optimum fit with respect to smoothness and data uncertainty. The repository uses an alternative form of the error function:

\begin{equation}
    0.5\cdot\left(1\pm\text{erf}\left[15\cdot\log_{10}\left(T/T_b\right)\right]\right)
    \label{eqn:erf_repo}
\end{equation}
where the factor $15$ is chosen for all fits to provide a reasonable half-width blending region (about $18 \%$ of $T_b$). The blend temperature $T_b$ is varied to give the minimum percent uncertainty. Figure \ref{fig:erf_func} shows the behavior of the error function and the blending operation of high and low temperature functions with exaggerated mismatch.

The repository offers various fitting functions for diverse applications. One physically-motivated example is the superconductivity function used for the low-temperature range of superconducting materials like aluminum (as described in Section \ref{sec:Al} below). An interpolation tool within the repository provides approximate fits between incomplete or separate datasets with large gaps in temperature coverage. 

\vspace{-1em}
\subsection{Compilation and Repository Structure}\label{sec:structure}

To simplify access to material fits, the repository generates compilation files named \verb+tc_compilation_<date>.csv+, which contain the details and parameters of a fit for each material. Table \ref{tab:comp_table} shows the format of a compilation file. Any fit can be reconstructed using the parameters given in the compilation file and the functions specified in the fit types file.

\begin{table}[ht]
    \centering
    \begin{tabular}{l|llllll}
        Material & Fit Type & $\text{T}_\text{low}$ & $\text{T}_\text{high}$ & \% Err & a & ...  \\ \hline
        Aluminum & Nppoly & 0.264 & 1.061 & 6.458 & -2.22 & ... \\ 
        BeCu & polylog & 2.0 & 80.0 & 2.0 & -0.50 & ... \\ 
        $\downarrow$ & $\downarrow$ & $\downarrow$ & $\downarrow$ &$\downarrow$ & $\downarrow$ & $\downarrow$  \\
    \end{tabular}
    \caption{The first rows of an example compilation file. Each row of the compilation file contains information about the fit type, range, and the fit parameters.} 
    \label{tab:comp_table}
    \vspace{-2em}
\end{table}

\begin{figure}[ht]
    \centering
    \includegraphics[width=\linewidth]{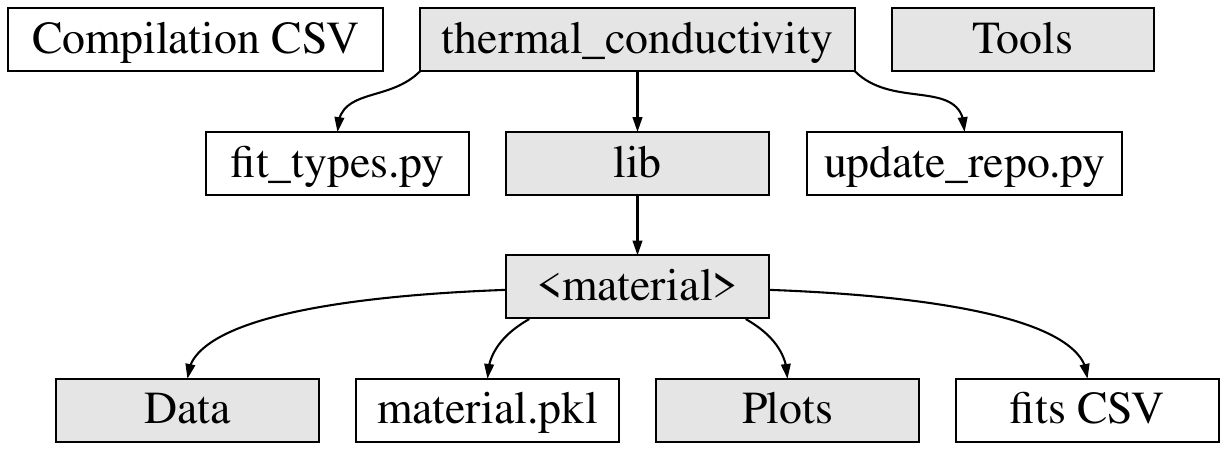}
    \caption{Simplified site map for the Cryogenic Materials Repository. Gray boxes indicate subdirectories, white boxes are notable files.}
    \label{fig:sit_map}
    \vspace{-0.75em}
\end{figure}

Within the \verb+thermal conductivity\+ subdirectory are the various Python files used to fit, organize, and plot the data. The \verb+lib\+ subdirectory houses a folder for each material and component stored in the repository. Each of these folders in turn stores the thermal conductivity data (when available), a compilation of the fit functions (sometimes multiple), and plots. The serialized material object, or instance of the Python \verb+Material+ class, is stored as a pickle file. Figure \ref{fig:sit_map} contains a simplified site map. 

\vspace{-1em}
\subsection{Tools}\label{sec:tools}

The CMR can be coupled with other codes to perform thermal loading calculations and modeling. The provided thermal conductivity fits can be integrated over a temperature range to yield loading estimates on cryogenic stages or predict thermal gradients. The repository contains a Python-based graphical user interface (GUI) tool to help in the development of a cryostat thermal model. The GUI includes built-in functions to perform cryogenic calculations, has liquid helium cryostat optimization functionality, and can easily be integrated with other Python-based codes. An example thermal model for a Bluefors LD400 system is included to serve as a reference.

\section{Thermal Conductivity Testing}\label{sec:tc_testing}
To expand the offerings of the CMR and to support the development of the next generation of sub-Kelvin systems, thermal conductivity measurements are being conducted at UT Austin. Thermal conductivity, or the ability of a material to conduct heat, is given by the equation:

\begin{equation}
    \dot{Q} = \frac{A}{\Delta x}\int_{T_1}^{T_2}\kappa(T)\mathrm{dT},
    \label{eqn:Qdot}
\end{equation}
where $\Delta x$ and $A$ are the length and cross-sectional area of the material, respectively. $\dot Q$ is the power applied to the material, and $\mathrm{dT}$ is the temperature difference across the material length. Equation \ref{eqn:Qdot} is often approximated by replacing the $\mathrm{dT}$ with a $\Delta T$ as shown in (\ref{eqn:k_approx}). The method described in \cite{hust_comments_1982} shows that this approximation is valid for $\Delta T$ values up to a similar magnitude of the bath temperature for temperature ranges that do not exhibit large changes in slope in conductivity. 

\begin{equation}\label{eqn:k_approx}
    \kappa(\bar T) = \frac{\Delta x}{A}\frac{\dot Q}{\Delta T}
\end{equation}

This work presents thermal conductivity testing for two varieties of CFRP and three different alloys of Aluminum. The specific setups, analysis methods, and results for each material are described in the respective sections below.

\subsection{Testing Setup}
To test below $1$ K, the UT Austin group uses a Bluefors LD400 Dilution Refrigerator (DR) with a minimum temperature ${\sim}7$ mK and ${\sim}400\;\mu$W of cooling capacity at $100$ mK. 
Test samples are mounted to the mixing chamber stage (MXC) with a resistive heater attached to the end of the sample furthest from the MXC. The heater is an Ohmite 89 Series Metal-Mite resistor chosen with either 50 $\Omega$, 100 $\Omega$, or 500 $\Omega$, depending on the desired power output. The heater resistances are stable to within ${\sim}2\%$ at cryogenic temperatures. The heater is connected to a Keithley 2280S Power Supply. Two Lake Shore Cryotronics R102A and R202A ruthenium oxide (ROX) thermometers calibrated between $0.05-40$ K are bolted to the component. Lake Shore specifies the systematic uncertainty of the ROXs as $\pm2$ mK. The R102A (T2) is placed near the heater, and the other (T1) near the coupling to the MXC such that $T_1 < T_2$. The mixing chamber stage can be operated using a proportional–integral–derivative (PID) controller to maintain a target bath temperature. 

The full temperature range is sampled by varying both the power supplied to the component heater and the MXC bath temperature. The combination of PID and power settings, referred to as setpoints, is determined by choosing values from a simulated thermal conductivity curve. The simulated curve can be extrapolated from a template curve in the repository or determined by fitting data from a similar material. A set of ${\sim}100$ evenly spaced points is chosen along the curve, and (\ref{eqn:Qdot}) is then inverted to determine the corresponding input power and MXC PID settings. These voltage and PID setpoints are then exported and serve as the configuration file for the testing sequence. 

During testing, periods of constant current are measured continuously by the PSU and will be referred to herein as frames. The control of the power supply, MXC PID, and time-ordered data (TOD) collection is completely automated using the Observatory Control System (OCS) \cite{koopman_simons_2024}.

\vspace{-0.75em}
\subsection{Analysis and Uncertainty}\label{sec:analysis}

To calculate the thermal conductivity, we first determine the power driven to the test component, $\dot Q$:
\begin{equation}
    \dot Q = I^2R_h,
\end{equation}
where $I$ is the average current of each frame, and $R_h$ is sample heater resistance. The standard deviation of the mean of the current is consistently subdominant to the manufacturer-defined systematic uncertainty of $\pm(0.05\% + 0.01 \text{ mA})$. 

The temperature of each thermometer is the steady-state value from the TOD in each frame, determined once five consecutive measurements are within the noise level of the thermometers (${\sim} 0.1$ mK). The change in temperature throughout the window must also remain within this threshold. The five-measurement window is evaluated at the end of the frame and expanded until these conditions are no longer met. The steady-state temperature and uncertainty are taken from the mean and standard deviation of the mean for all measurements in the final window.

In the event the thermometer does not reach steady-state by the end of a frame, the temperature data is fit using the following asymptotic regression function: 
\begin{equation}
    T(t) = A - (A-B)\mathrm{e}^{-Ct},
    \label{eqn:asymptotic_regression}
\end{equation}
where $A$ is the asymptotic steady-state temperature value of the TOD. The uncertainty of this measurement is found from the covariance matrix of the asymptotic fit. 

A number of cleaning cuts are performed on the data. First, frames with unstable currents are removed. Next, frames in which the thermometer temperatures cannot be determined using either of the above methods are removed. Last, frames during which the mixing chamber was unstable are removed. Altogether, these cuts remove ${\sim} 5\%$ of the data points. 

The parasitic power load from the thermometer and heater wires is determined by fitting the power against each thermometer temperature with a quadratic function. The point of convergence of these quadratic fits indicates the power that would achieve a $\Delta T = 0$, which is used as the parasitic heat for the run. 

Lastly, (\ref{eqn:k_approx}) is used to determine the thermal conductivity from each of these values derived from the TOD.

\vspace{-1em}
\subsection{Carbon Fiber Reinforced Polymers}
The two CFRP samples tested were the 7:8 mm ID:OD CFRP tubes from \textit{DPP Pultrusion}, and the 0.25:0.32 inch ID:OD CFRP tubes from \textit{Clearwater Composites LLC}. The mechanical capabilities of tubes by these manufacturers were tested by Crowley et al., but the collected thermal data covered a more limited range, motivating follow-up measurements \cite{crowley_simons_2022}. The \textit{Clearwater} tubes are roll-wrapped with a twill fabric exterior. The \textit{DPP} tubes are created via an axial pultrusion process. 

\begin{figure*}[htbp] 
    \centering
    \includegraphics[width=0.80\textwidth]{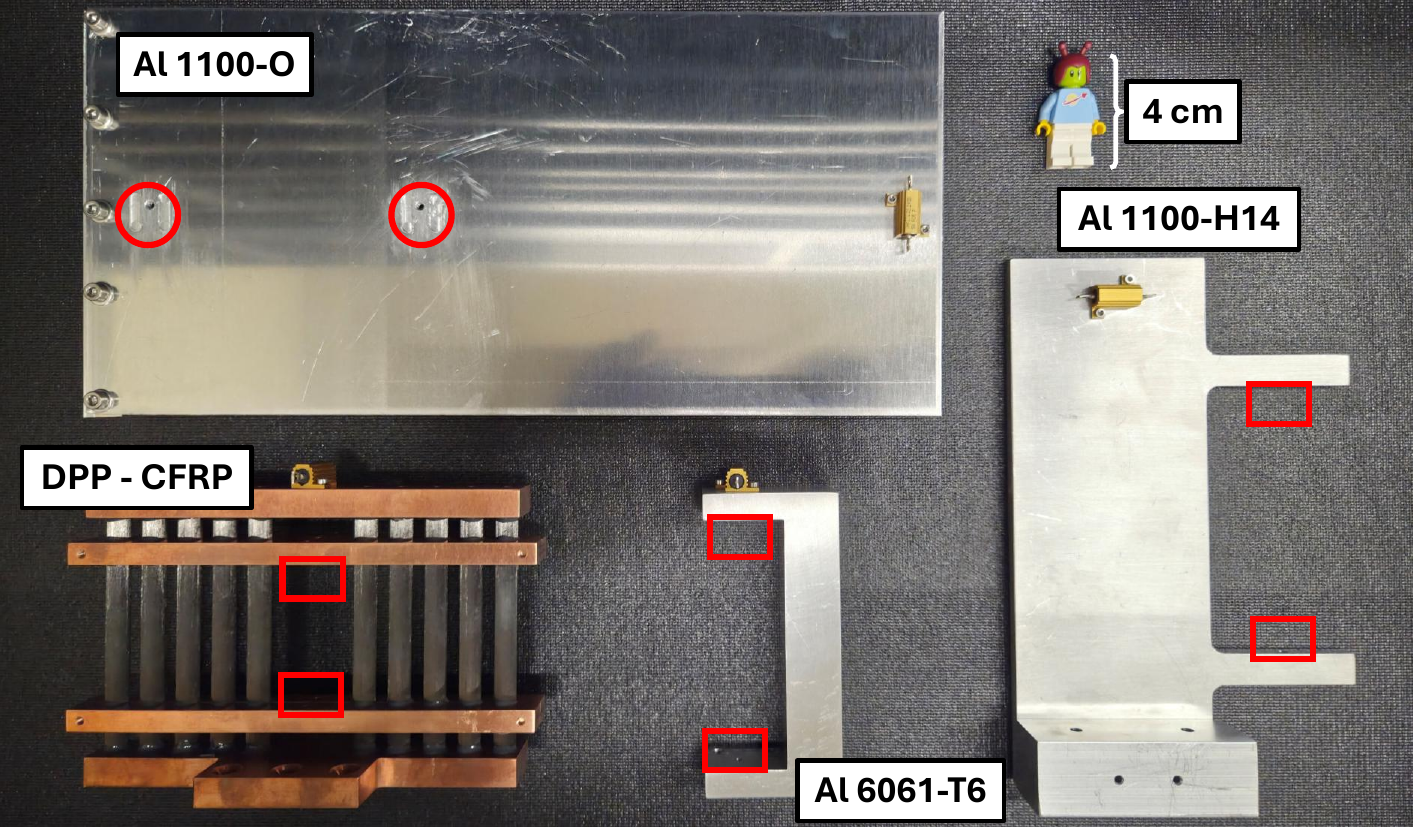}
    \caption{Tested aluminum and the \textit{DPP} carbon fiber samples. The positioning of the ROX thermometers is indicated with red circles/rectangles. The \textit{Clearwater} and \textit{DPP} CFRP test samples are identical in build. A $4$ cm mini figure is given for scale. The aluminum dimensions are given in Table \ref{tab:Al-comp}.}
    \label{fig:SampleZoo}
    \vspace{-1em}
\end{figure*}

Figure \ref{fig:SampleZoo} shows the testing setup for the \textit{DPP} CFRP tubes. The apparatus holding the CFRP tubes was composed of high thermal conductivity oxygen-free high-conductivity copper. Both ends of each tube were fixed within a machined hole in the copper using epoxy. The test heater of $50$ $\Omega$ was mounted to the bottom copper bar. Each of the test thermometers was attached to separate copper cross pieces to minimize any non-uniform thermal gradients from the test and MXC stage heaters, and to minimize the uncertainty in the vertical distance between the two thermometers. The test apparatus for the \textit{Clearwater} tubes followed the same design.

An additional $10\text{ k}\Omega$ resistor was added to the heater circuit in series to increase the total resistance of the circuit and allow the power supply to operate further from its specified limits. The $10\text{ k}\Omega$ resistor was clamped to the $1$ K DR stage. Between the \textit{DPP} and \textit{Clearwater} tests, additional heatsinking of the thermometer and heater wires changed the parasitic load from $29.4 \pm 9.0 \text{ nW}$ to $17 \pm 6.3 \text{ nW}$.

\vspace{-1em}
\subsection{Aluminum Alloys}\label{sec:Al}

The three different aluminum alloys tested were Aluminum 1100-O, 1100-H14, and 6061-T6. Each alloy designation indicates a different composition, which influences the mechanical and thermal properties of the material. The specific elemental compositions of the components are shown in Table \ref{tab:Al-comp}. Each sample was machined to minimize non-uniform thermal gradients at the thermometers and minimize the uncertainty in the distance between the thermometers. Due to the differing thickness of the aluminum material used, each sample test geometry is unique. Figure \ref{fig:SampleZoo} shows each aluminum component. The dimensions of each sample are outlined in Table \ref{tab:Al-comp}.



\begin{table}[]
\centering
\begin{tabular}{r|p{1.75cm}p{1.75cm}p{2.0cm}}
Material & Al 6061-T6 & Al 1100-O & Al 1100-H14 \\
Source & Castle Metals Aerospace & Trinity Brand Industries & Pierce Aluminum Co \\ 
(mm) & 89.8$\times$20.3$\times$25.4 & 100$\times$152$\times$3.175 & 100.4$\times$73$\times$6.35\\\hline
Al & 97.40 & 99.00 & 99.00 \\
Mg & 0.90 & -- & -- \\
Si & 0.75 & 0.16 & \multirow{2}{*}{Fe+Si = 0.95} \\
Fe & 0.38 & 0.38 &  \\
Cu & 0.31 & 0.07 & 0.05-0.20 \\
\multicolumn{4}{c}{Other elements all $< 0.1\%$}\\\end{tabular}
\caption{Aluminum Alloy composition details with elemental abundances shown in percent. Sample dimensions, reported as $\Delta$L$\times$W$\times$T, describe the cross-sectional area and distance between the thermometers.}
\label{tab:Al-comp}
\vspace{-2em}
\end{table}

For a superconducting material, the thermal conductivity is a function of the phonon contribution and the normal/superconducting electron contributions \cite{sauvage_new_2022}.
\begin{equation}\label{eqn:al_normal_kappa}
    \kappa_\text{normal} = aT^b + cT^b
\end{equation}
\begin{equation}\label{eqn:al_super_kappa}
    \kappa_\text{superconducting} = \alpha T^\beta + \gamma T e^{\delta/T}
\end{equation}

Combining these with (\ref{eqn:Qdot}), we obtain :
\begin{equation}
    \dot{Q}_\text{normal} = \frac{A}{x}\int_{T_1}^{T_2}\left(aT^b + cT^b\right)\mathrm{dT} + \dot Q_0
    \label{eqn:QdotsuperFit}
\end{equation}
\begin{equation}
    \dot{Q}_\text{superconducting} = \frac{A}{x}\int_{T_1}^{T_2}\left(\alpha T^\beta + \gamma T e^{\delta/T}\right)\mathrm{dT} + \dot Q_0
    \label{eqn:QdotnormalFit}
\end{equation}

In these expressions, the integrand is the $\kappa(T)$ function for thermal conductivity. Thus, the function parameters can be determined directly from the measured T, $\dot Q$, and geometric properties instead of fitting the $\kappa(\bar T)$ data, thereby not relying on the $\Delta T$ approximation. This is only made possible by having a physically motivated analytic form for the thermal conductivity of the material. The constant $\dot Q_0$ terms in the functions represent the parasitic power in the system, which can be directly estimated via the fit parameters.

\vspace{-0.5em}
\section{Results}\label{sec:results}

\subsection{Carbon Fiber Reinforced Polymers}

\begin{figure}[ht]
    \centering
    \includegraphics[width=0.96\linewidth, trim={10 10 0 10}]{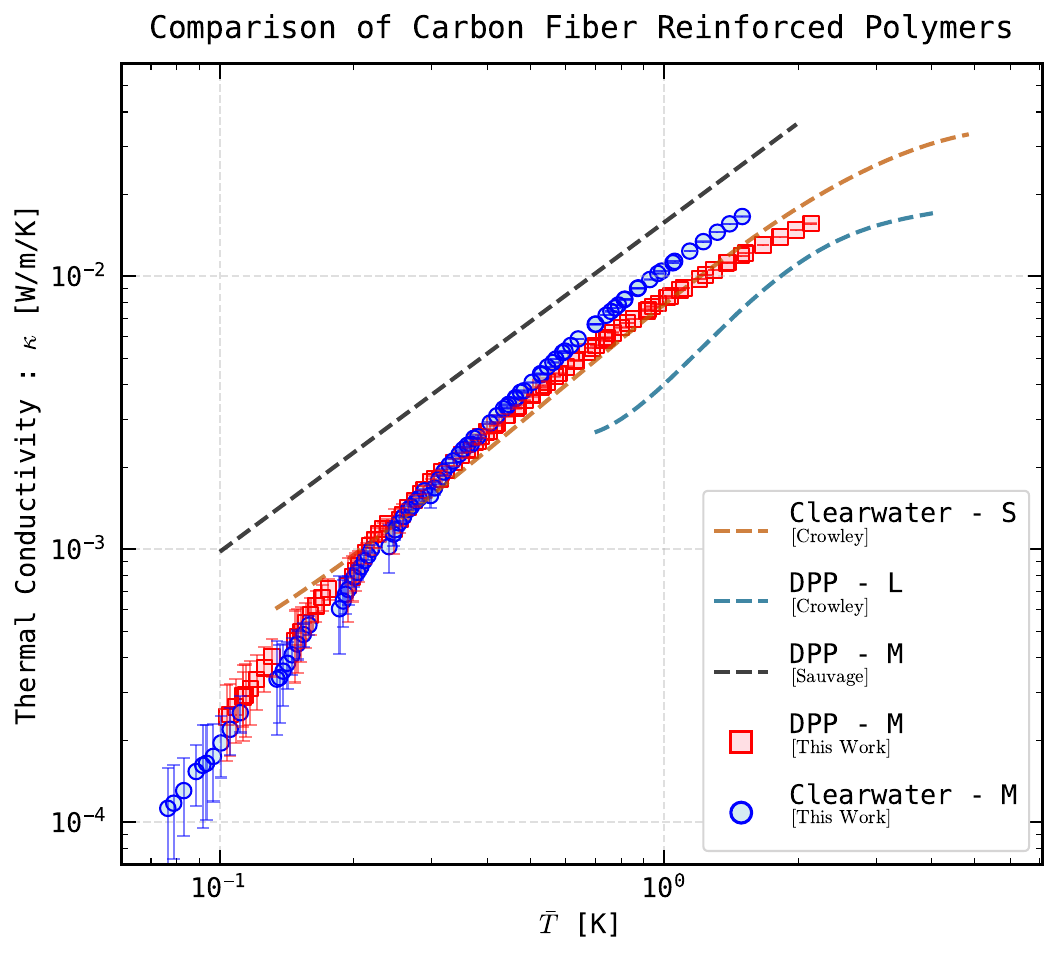}
    \caption{Thermal conductivity of CFRP. \textit{Clearwater} Small (S) measures 1.68:2.06 mm (ID:OD), \textit{DPP} Large (L) is a 10 mm solid rod, \textit{DPP} Medium (M) measures 6.8:7.8 mm (ID:OD), and \textit{Clearwater} Medium (M) is 0.25:0.32 in (ID:OD). Shaded data points represent this work, dashed lines are fits to previous measurements.}
    \label{fig:CFRP_test}
    \vspace{-2em}
\end{figure}

Figure \ref{fig:CFRP_test} shows the results of the thermal conductivity testing of the \textit{Clearwater} and \textit{DPP} CFRP samples in context with previous measurements. The \textit{Clearwater} sample results agree with previous measurements of \textit{Clearwater} CFRP performed in \cite{crowley_simons_2022}. The \textit{DPP} results lie between the range of previous results for \textit{DPP} in \cite{crowley_simons_2022} and \cite{sauvage_new_2022}. Variation from the values measured for \textit{DPP} previously is currently unexplained but may be due to differences in the specific component dimensions, testing setups, or more. These results do not indicate any significant difference in the thermal conductivities of tubes from the two different manufacturers. However, given the difference in cross-sectional area, the conductance of these commercially available \textit{Clearwater} tubes is on average $(2.12\pm 0.25)\times$ that of the \textit{DPP} tubes.

\vspace{-1em}
\subsection{Aluminum Alloys}
To obtain the thermal conductivity fit parameters for the three aluminum alloys, the power, $\dot Q$, is fit as a function of the average temperature of the thermometers, $\bar T$. The fitting is performed using the scipy \verb+optimize+ function and the functional forms defined in equations \ref{eqn:QdotsuperFit} and \ref{eqn:QdotnormalFit}. 
The transition temperature between the two fit functions is $1.2$ K, the superconducting transition temperature of the aluminum alloy \cite{rose-innes_introduction_1978}.

\begin{figure}[ht]
\vspace{-0.5em}
    \centering
    \includegraphics[width=0.95\linewidth, trim={10 10 10 10}]{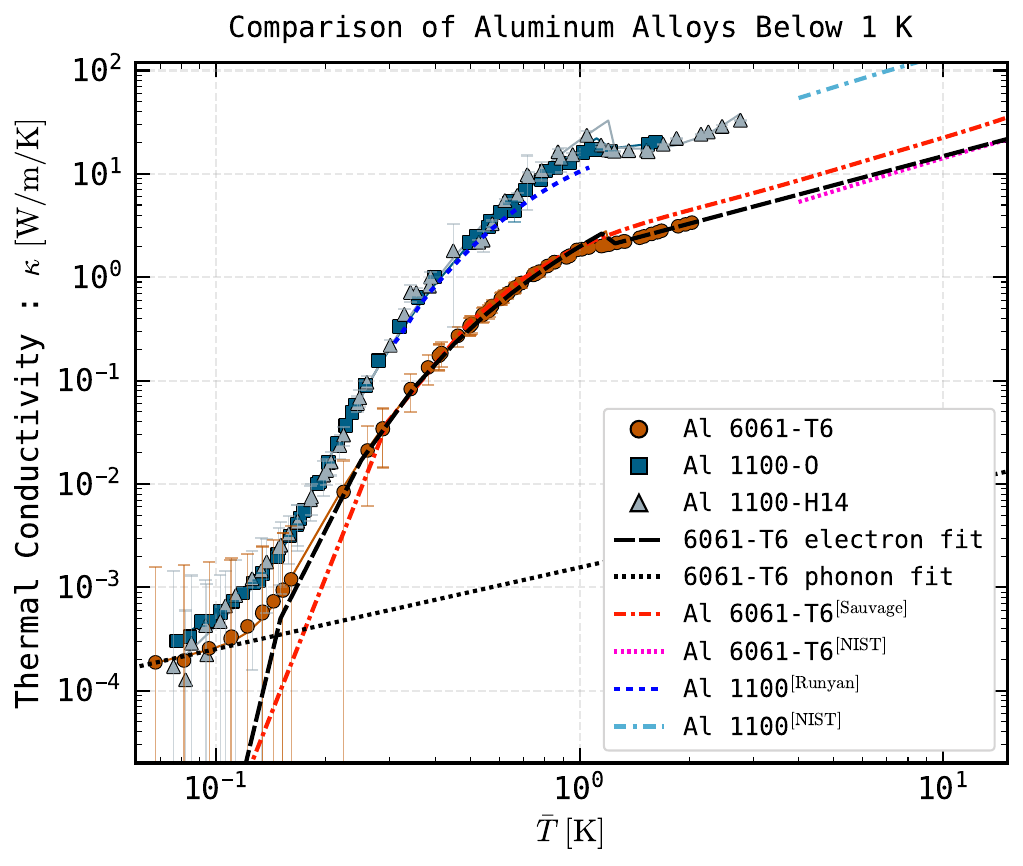}
    \vspace{-0.5em}
    \caption{Thermal conductivity of aluminum alloys with fits accounting for parasitic heat. The separated phonon and electron contribution curves are shown for Al 6061-T6 in dotted and dashed black lines, respectively.}
    \label{fig:al_alloy_plot}
\end{figure}

Figure \ref{fig:al_alloy_plot} shows the thermal conductivity determined using the same method described above for CFRP. Each fit is of the form shown in equations \ref{eqn:QdotsuperFit} and \ref{eqn:QdotnormalFit}, and the fit parameters are listed in Table \ref{tab:al_fit_params}. With the physically-motivated functional form, the thermal conductivity fits for each material can be decomposed into phonon and electron contribution curves, also shown in figure \ref{fig:al_alloy_plot}. At lower temperatures, the total thermal conductivity is dominated by the phonon contribution, and the aluminum curves begin to flatten out. This is supported by previous measurements and theory shown in \cite{ohara_resonant_1974} and \cite{ventura_thermal_1998}. However, large uncertainties at the coldest temperatures motivate further investigation of this phenomenon. The phonon contribution curve is particularly useful for extrapolating to even lower temperatures.

Here, the measurements are shown to be consistent with fits at temperatures ${\geq}4$ K from the NIST website. These measurements agree with previous results in \cite{runyan_thermal_2008}, and \cite{sauvage_new_2022}.

\begin{table}[h]
    \centering
    \begin{tabular}{l|ccc}
     & Al 6061-T6 & Al 1100-O & Al 1100-H14 \\
    \hline
    a/$\alpha$ & 2.14e-03 & 4.50e-03 & 9.94e-01 \\
    b/$\beta$ & 9.34e-01 & 9.93e-01 & 3.37e+00 \\
    $\gamma$ & 6.39e+00 & 6.31e+01 & 8.86e+01 \\
    $\delta$ & -1.16e+00 & -1.35e+00 & -1.48e+00 \\
    $c$ & 1.70e+00 & 1.62e+01 & 1.98e+01 \\
    $d$ & 9.60e-01 & 4.16e-01 & -9.43e-01 \\ \hline
    T$_\text{min}$ & 6.78e-02 & 7.38e-02 & 7.61e-02 \\
    T$_\text{max}$ & 2.03e+00 & 1.62e+00 & 2.77e+00 \\
    \end{tabular}
    \caption{Thermal conductivity fit parameters corresponding to functions \ref{eqn:al_normal_kappa} and \ref{eqn:al_super_kappa} for each of the tested aluminum alloys.}
    \label{tab:al_fit_params}
\end{table}

\vspace{-2.5em}
\subsection{Material Comparison}
\begin{figure}
    \centering
    \includegraphics[width=0.96\linewidth, trim={0 10 0 0}]{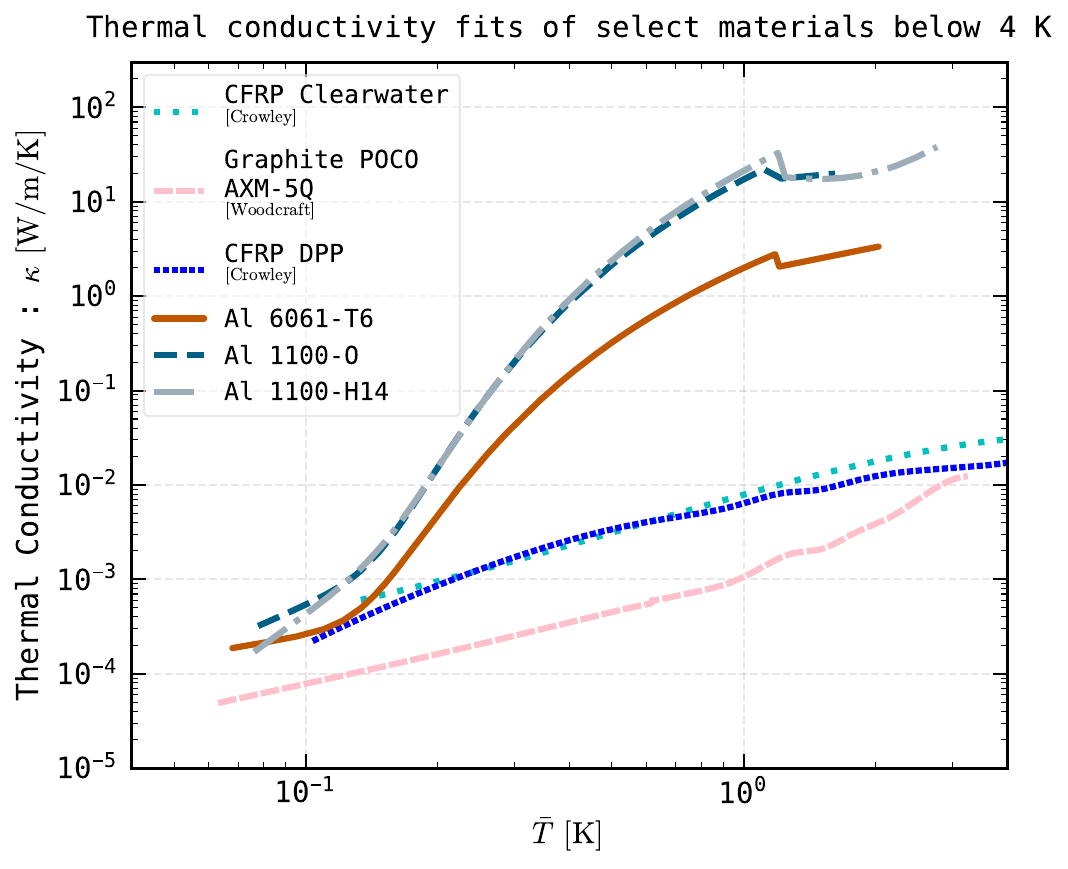}
    \vspace{-0.75em}
    \caption{Thermal conductivity fits of aluminum alloys, carbon fiber, and graphite varieties. Thermal conductivities appear to converge around $100 \text{ mK}$, but more data at even colder temperatures would further distinguish the properties of these materials.}
    \label{fig:low_temp_materials}
    \vspace{-1em}
\end{figure}

Figure \ref{fig:low_temp_materials} compares the thermal conductivity of CFRP, Aluminum, and Graphite at temperatures below $\sim 3$K. These materials are of particular interest for the development of focal plane architectures at and below $100$ mK for the low temperature detector community \cite{crowley_simons_2022}. These results indicate that as temperatures decrease, the thermal conductivity of aluminum alloys levels off at a value similar to that of other available materials. This plot, and other similar comparison plots, are easily produced using the CMR. 

Large uncertainties for the Al 6061-T6 measurement are attributed to the systematic uncertainty in the single current measurement per frame which dominates at low currents. Later upgrades to the readout process enabled continuous current monitoring, decreasing its uncertainty in subsequent material testing. For all materials, constant parasitic power uncertainty is the largest contributor to uncertainty below ${\sim}100$ mK, with uncertainty at higher temperatures limited by the heater resistance systematic uncertainty. Measurements below ${\sim}100$ mK were limited by the parasitic load from the necessary cables and the minimum current restrictions of the power supply. Planned improvements by decreasing the parasitic load with better heatsinking, decreasing the thermometer calibration, and increasing the sample $A/L$ will further enable measurements at the coldest temperatures. 
\vspace{-0.5em}
\section{Conclusion}
In summary, this work presents the CMR, a public, Python-based repository for storing material properties and fits across a wide temperature range. The repository stores thermal conductivity data and published fits in an accessible format. It employs a transparent and adaptable fitting algorithm to create new fits from that data. The repository is currently accessible as a public GitHub repository, enabling adaptation and updates to meet the community's future needs. Planned developments will include additional material properties, data, materials, and custom components. 

Sub-kelvin thermal conductivity tests conducted at UT Austin are expanding the offerings of the CMR database. Measurement results of the CFRP samples are in agreement with previous measurements, and so far do not show a significant difference between different manufacturers. The results from Al 6061-T6, 1100-O, and 1100-H14 are also in agreement with previous measurements. Our measurements augment the data available at sub-kelvin temperatures and highlight the distinction between aluminum alloys. Continued testing of samples of varying materials and components will further expand the available data on cryogenic material properties. This repository and the newest material testing will serve as a valuable tool to the low-temperature detector community.

\vspace{-0.25em}
\section*{Acknowledgment}
Special thanks to J. Groh, B. Jones, J. Olson, and others for their sharing of data and collaboration. To K. Rink and A. Salvarese for their contributions. And to P. Bradley, V. Sauvage, and R. Snodgrass for their feedback and consultation.

\bibliographystyle{IEEEtran}
\bibliography{LTD2025}{}

\newpage


\begin{IEEEbiographynophoto}{Henry E. Nachman}
was born in Chapel Hill, NC, USA in 2001. He received the B.S. degree in astrophysics from The University of North Carolina at Chapel Hill, Chapel Hill, NC, USA, in 2023. 

Since 2023, he is a graduate student at The University of Texas at Austin and a member of the Weinberg Institute for Theoretical
Physics, Texas Center for Cosmology and Astroparticle Physics, Austin, TX, USA.
\end{IEEEbiographynophoto}
\begin{IEEEbiographynophoto}{Ray Radebaugh} was born in South Bend, IN, USA, in 1939. He received the B.S.
degree in engineering physics from the University of Michigan, Ann Arbor, MI, USA, in
1962, the M.S. degree in physics from Purdue University, West Lafayette, IN, USA, in
1965, and the Ph.D. degree in physics from Purdue University, West Lafayette, IN, USA,
in 1966.

From 1966 to 1968, he was a postdoctoral fellow with the National Bureau of
Standards, Boulder, CO, USA. From 1968 to 1995 he was a physicist with the National
Bureau of Standards/National Institute of Standards and Technology, Boulder, CO, USA.
From 1995 to 2009 he was the leader of the Cryogenic Technologies Group of the
National Institute of Standards and Technology (NIST), Boulder, CO, USA. His research
was primarily in the field of cryogenic refrigeration and cryogenic material properties.
He retired in 2009 as a NIST Fellow Emeritus.

Dr. Radebaugh was on the board of directors for the International Cryocooler
Conference and had more than 200 publications. He passed away in Aurora, CO, USA, on February 11, 2026.
\end{IEEEbiographynophoto}

\begin{IEEEbiographynophoto}{Oorie Desai, Nicholas Galitzki, Daniel Lee, JB Lloyd, Tannishtha Nandi, Ani
Pagni, and Elle C. Shaw} 
author biographies not available at the time of publication.
\end{IEEEbiographynophoto}

\vfill

\end{document}